\documentclass[aps,prl,twocolumn,showpacs,amsmath,amssymb,final,letterpaper]{revtex4-1}
\usepackage{graphicx}
\usepackage{subfigure}
\usepackage{txfonts}

\begin{document}

\title{Structural preferential attachment: Network organization beyond the link.}

\author{Laurent H\'ebert-Dufresne}
\author{Antoine Allard}
\author{Vincent Marceau}
\author{Pierre-Andr{\'e} No\"el}
\author{Louis J. Dub\'e}

\affiliation{D\'epartement de Physique, de G\'enie Physique, et d'Optique, Universit\'e Laval, Qu\'ebec (Qu{\'e}bec), Canada G1V 0A6}
\date{\today}
\begin{abstract}
We introduce a mechanism which models the emergence of the universal properties of complex networks, such as \emph{scale independence}, \emph{modularity} and \emph{self-similarity}, and unifies them under a scale-free organization beyond the link. This brings a new perspective on network organization where communities, instead of links, are the fundamental building blocks of complex systems. We show how our simple model can reproduce social and information networks by predicting their community structure and more importantly, how their nodes or communities are interconnected, often in a self-similar manner.
\end{abstract}
\pacs{89.75.Da, 89.75.Fb, 89.75.Hc, 89.75.Kd, 89.65.Ef}
\maketitle

\paragraph*{\textbf{A universal matter.}}
Reducing complex systems to their \emph{simplest possible form} while retaining their important properties helps model their behavior independently of their nature. Results obtained via these abstract models can then be transferred to other systems sharing a similar simplest form. Such groups of analog systems are called \emph{universality classes} and are the reason why some models apply just as well to the sizes of earthquakes or solar flares than to the sales number of books or music recordings \cite{newmanPL}. That is, their statistical distributions can be reproduced by the same mechanism: \emph{preferential attachment}. This mechanism has been of special interest to network science \cite{barabasi09} because it models the emergence of power-law distributions for the number of links per node. This particular feature is one of the universal properties of network structure \cite{barabasi99}, alongside modularity \cite{newman02} and self-similarity \cite{song}. Previous studies have focused on those properties one at a time \cite{barabasi99,newman02,song,barabasi00,guimera2,song06}, yet a unified point of view is still wanting. In this Letter, we present an overarching model of preferential attachment that unifies the universal properties of network organization under a single principle.

Preferential attachment is one of the most ubiquitous mechanisms describing how elements are distributed within complex systems. More precisely, it predicts the emergence of \emph{scale-free} (power-law) distributions where the probability $P_k$ of occurrence of an event of order $k$ decreases as an inverse power of $k$ (i.e., $P_k \propto k^{-\gamma}$ with $\gamma > 0$). It was initially introduced outside the realm of network science by Yule \cite{yule} as a mathematical model of evolution explaining the power-law distribution of biological genera by number of species. Independently, Gibrat \cite{gibrat} formulated a similar idea as a law governing the growth rate of incomes. Gibrat's law is the sole assumption behind preferential attachment: the growth rates of entities in a system are proportional to their size. Yet, preferential attachment is perhaps better described using Simon's general \emph{balls-in-bins} process \cite{simon}.

\begin{figure*}[!htb]
  \centering
  \subfigure[]{\includegraphics[width=0.3\linewidth]{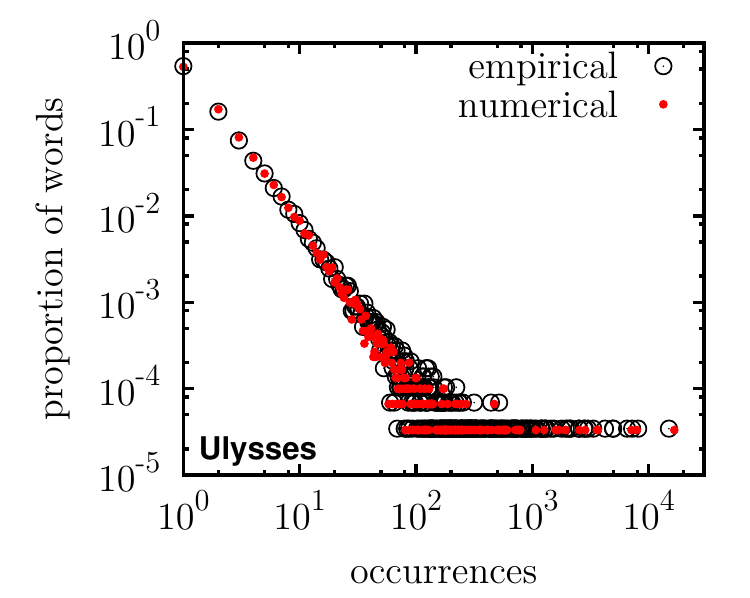} \label{ulysse}}     
  \subfigure[]{\includegraphics[width=0.3\linewidth]{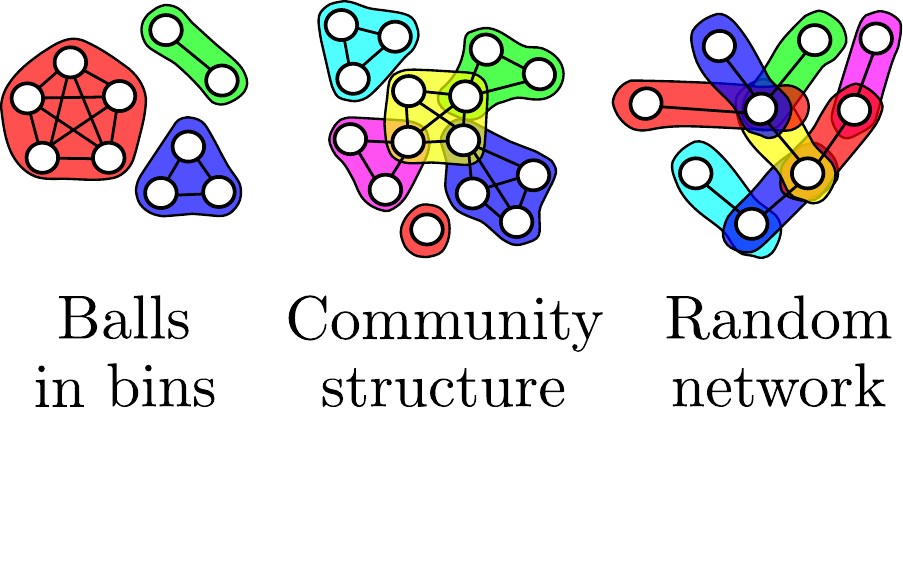} \label{strucs}}       
  \subfigure[]{\includegraphics[width=0.3\linewidth]{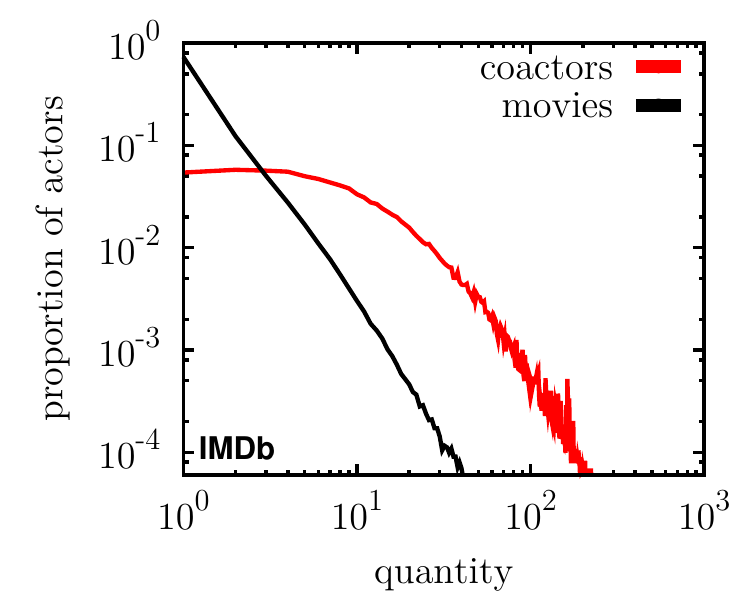} \label{imdb}}
  \caption{(Color online) (a) The distribution of words by their number of appearances in James Joyce's Ulysses (empirical data). The numerical data was obtained from a single realization of Simon's model with $p$ equal to the ratio of unique words (30 030) on the total word count (267 350). (b) Schematization of the systems considered in this Letter, illustrating how order (Simon's model of balls in bins) and randomness (Barab\'{a}si-Albert's model of random networks) coexist in a spectrum of complex systems. (c) The distribution of coactors and movies per actor in the Internet Movie Database since 2000. The organization moves closer to a true power law when looking at a higher structural level (i.e., movies versus coactors).}
  \label{FIG1}
\end{figure*}

Simon's model was developed for the distribution of words by their frequency of occurrence in a prose sample \cite{zipf}. The problem is the following: what is the probability $P_{k+1}(i+1)$ that the $(i+1)$\emph{-th} word of a text is a word that has already appeared $k$ times? By simply stating that $P_{k+1}(i+1) \propto k\cdot P_k(i)$, Simon obtained the desired distribution [Fig. \ref{ulysse}]. In this model, the nature of the system is hidden behind a simple logic: the ``popularity'' of an event is encoded in its number of past occurrences. More clearly, a word used twice is 2 times more likely to reappear next than a word used once. However, before its initial occurrence, a word has appeared exactly zero times, yet it has a certain probability $p$ of appearing for the very first time. Simon's model thus produces systems whose distribution of elements falls as a power law of exponent $\gamma = (2-p)/(1-p)$.

\paragraph*{\textbf{On the matter of networks.}}
Networks are ensembles of potentially linked elements called \emph{nodes}. In the late 1990s, it was found that the distribution of links per node (the \emph{degree distribution}) featured a power-law tail for networks of diverse nature. To model these so-called \emph{scale-free networks}, Barab\'{a}si and Albert \cite{barabasi99} introduced preferential attachment in network science. In their model, nodes are added to the network and linked to a certain number of existing nodes. The probability that the new node chooses an old one of degree $k$ is proportional to $k\cdot N_k$, where $N_k$ is the number of nodes of degree $k$. As the system goes to infinity, $N_k$ falls off as $k^{-3}$.

From the perspective of complex networks, Simon's model may be regarded not as a scheme of throwing balls (e.g., word occurrences) in bins (e.g., unique words), but as an extreme case of scale-free networks where all links are shared within clearly divided structures. Obviously, both Simon's and the Barab\'{a}si-Albert's (BA) models follow the preferential attachment principle. However, Simon's model creates distinct growing structures, whereas the BA model creates overlapping links of fixed size. By using the same principle, one creates order while the other creates randomness [Fig.\ref{strucs}]. Our approach explores the systems that lie in between.

\paragraph*{\textbf{When structure matters.}}
The vast majority of natural networks have a \textit{modular topology} where links are shared within dense subunits \cite{newman02}. These structures, or \emph{communities}, can be identified as social groups, industrial sectors, protein complexes or even semantic fields \cite{palla05}. They typically overlap with each other by sharing nodes and their number of neighboring structures is called their \emph{community degree}. This particular topology is often referred to as \emph{community structure} [Fig. \ref{strucs}]. Because these structures are so important on a global level, they must influence local growth. Consequently, they are at the core of our model.

The use of preferential attachment at a higher structural level is motivated by three observations. First, the number of communities an element belongs to, its \emph{membership} number, is often a better indicator of its activity level than its total degree. For instance, we judge an actor taking part in many small dramas more active than one cast in a single epic movie as one of a thousand extras, as we may consider a protein part of many complexes more functional than one found in a single big complex.

Second, studies have hinted that Gibrat's law holds true for communities within social networks \cite{rybski}. The power-law distribution of community sizes recently observed in many systems (e.g., protein interaction, word association and social networks \cite{palla05} or metabolite and mobile phone networks \cite{ahn}) supports this hypothesis. 

Third, degree distributions can deviate significantly from true power laws, while higher structural levels might be better suited for preferential attachment models [Fig. \ref{imdb}].% This idea is strengthened by evidence of \textit{self-similar organization} in complex networks \cite{song}, indicating that under certain renormalization procedures one can transform community structured systems into scale-free random networks.

\paragraph*{\textbf{A simple model.}}
Simon's model assigns elements to structures chosen proportionally to their sizes, while the BA model creates links between elements chosen proportionally to their degree. We thus define \emph{structural preferential attachment} (SPA), where both elements and structures are chosen according to preferential attachment. Here, links will not be considered as a property of two given nodes, but as part of structures that can grow on the underlying space of nodes and eventually overlap.

\begin{figure}[h!]
\centering
\includegraphics[width=0.9\linewidth]{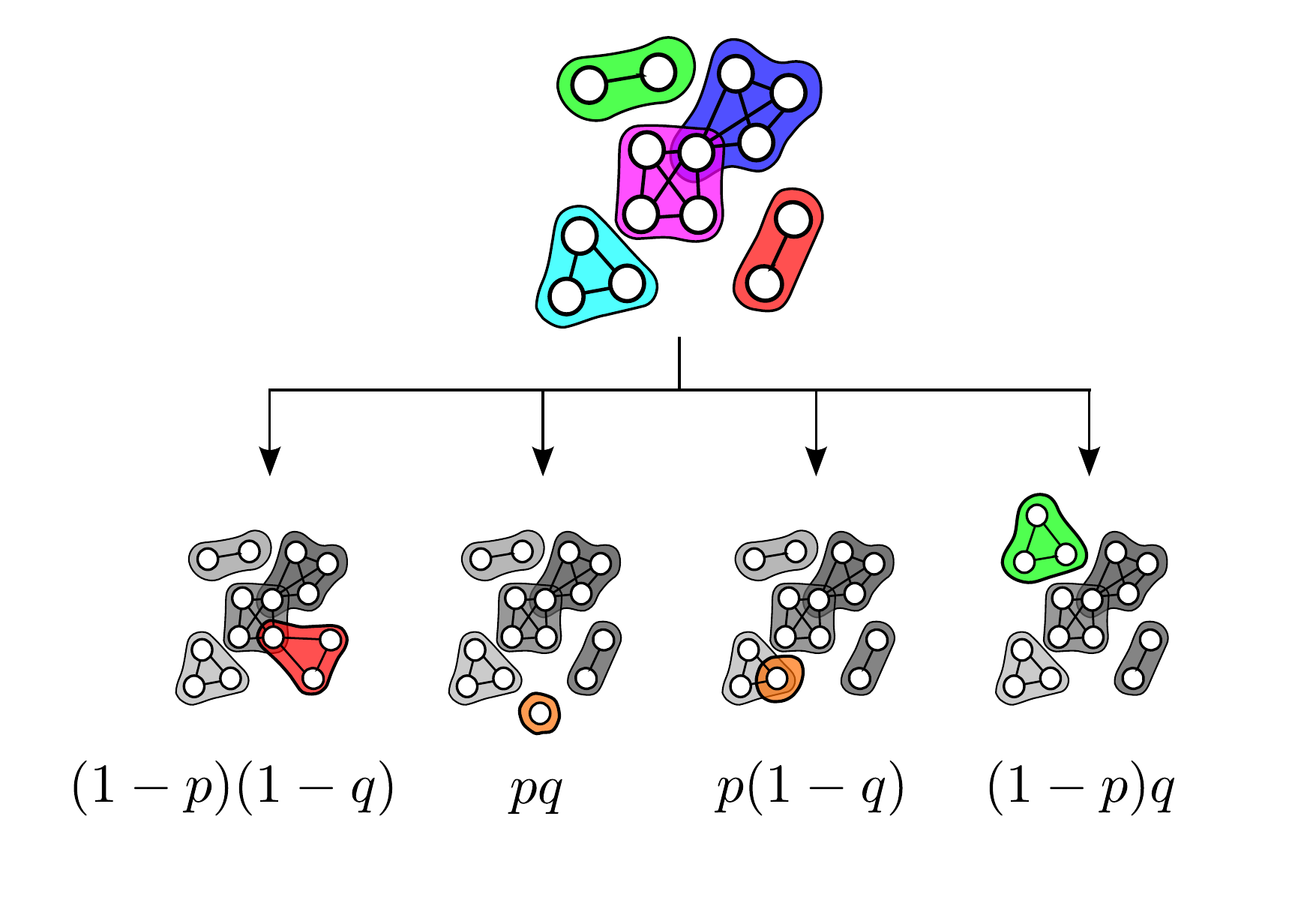}\\
\vspace{2.0\baselineskip}
\includegraphics[width=0.95\linewidth]{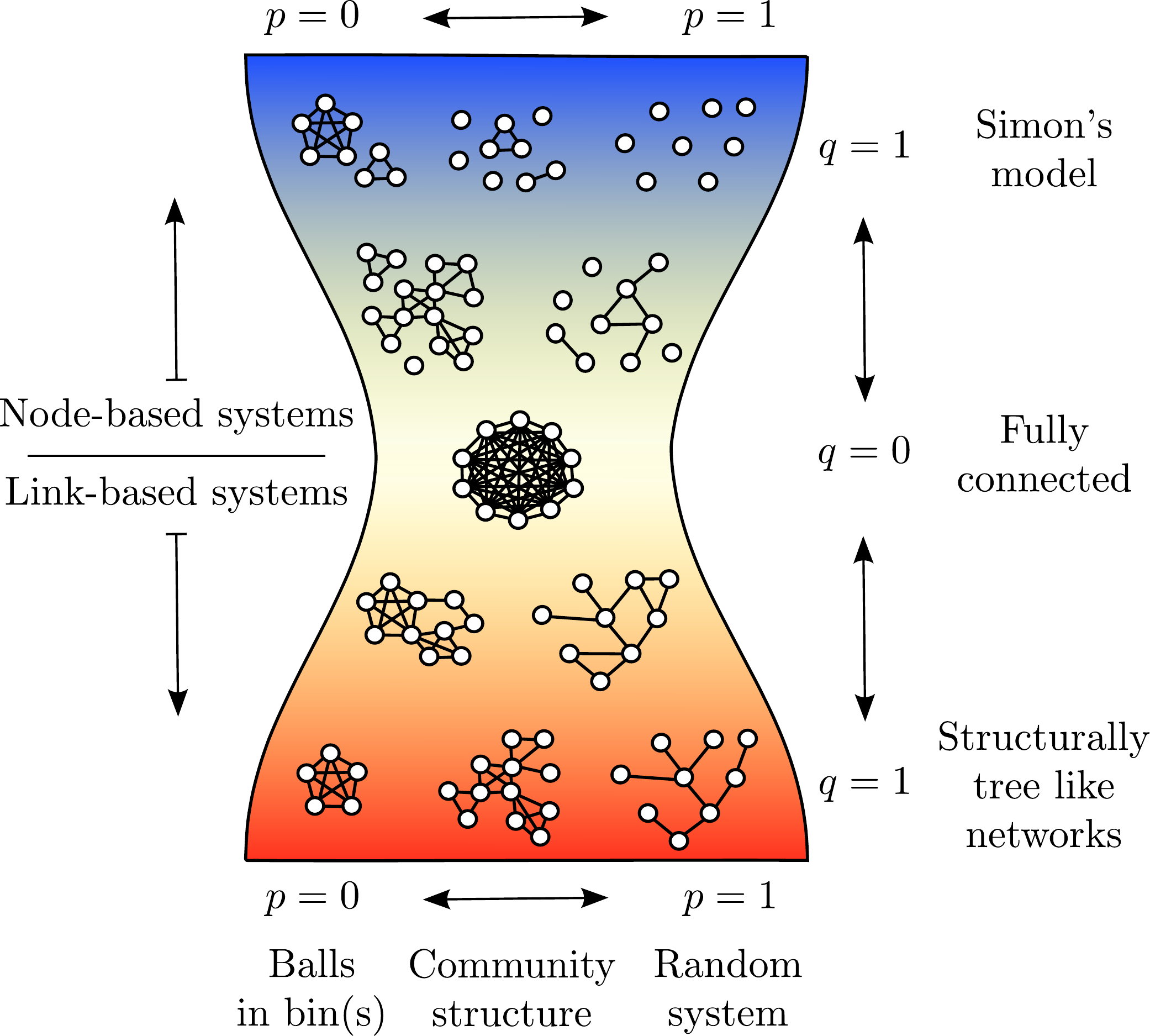}
\caption{(Color online) (top) Representation of the possible events in a step of node-based SPA; the probability of each event is indicated beneath it. (bottom) A schematization of the spectrum of systems obtainable with SPA. Here, we illustrate the conceptual differences between node-based $s=1$ and link-based systems $s=2$: Simon's model ($q=1$) creates structures of size one (nodes), while the BA model ($p=q=1$) creates random networks through structures of size two (links).}
\label{spectrum}
\end{figure}

Our model can be described as the following stochastic process. At every time step, a node joins a structure. The node is a new one with probability $q$, or an old one chosen proportionally to its membership number with probability $1-q$. Moreover, the structure is a new one of size $s$ with probability $p$, or an old one chosen among existing structures proportionally to their size with probability $1-p$. These two growth parameters are directly linked to two measurable properties: modularity ($p$) and connectedness ($q$)  [Fig. \ref{spectrum}]. Note that, at this point, no assumption is made on how nodes are linked within structures; our model focuses on the modular organization.

Whenever the structure is a new one, the remaining $s-1$ elements involved in its creation are once again preferentially chosen among existing nodes. The basic structure size $s$ is called the \emph{system base} and refers to the smallest structural unit of the system. It is not a parameter of the model \textit{per se}, but depends on the considered system. For instance, the BA model directly creates links, i.e. $s=2$ (with $p=q=1$), unlike Simon's model which uses $s=1$ (with $q=0$). All the results presented here use a node-based representation ($s=1$), although they can equally well be reproduced via a link-based representation ($s=2$). In fact, for sufficiently large systems, the distinction between the two versions seems mainly conceptual (see Supplemental Material for details \cite{SM}).

In our process, the growth of structures is not necessarily dependent on the growth of the network (i.e., the creation of nodes). Consequently, we can reproduce statistical properties of real networks without having to consider the large-size limit of the process. This allows our model to naturally include finite size effects (e.g., a distribution cutoff) and increases freedom in the scaling properties. In fact, we can follow $S_n$ and $N_m$, respectively, the number of structures of size $n$ and of nodes with $m$ memberships, by writing master equations for their time evolution \cite{evolution}:

\begin{equation}
\dot{S}_n(t) = ( 1\! -\! p )\frac{( n - 1)S_{n-1}(t) - nS_n(t)}{\left[1+p(s-1)\right]t} + p\delta _{n,s} \; ;
\label{m1}
\end{equation}
\begin{equation}
\dot{N}_m(t) = ( 1\! + \! p(s-1) - q )\frac{( m\! -\! 1)N_{m-1}(t) - mN_m(t)}{\left[1+p(s-1)\right]t} + q\delta _{m,1} \; .
\label{m2}
\end{equation}
Equations (\ref{m1}) and (\ref{m2}) can be transformed into ODEs for the evolution of the distribution of nodes per structure and structure per node by normalizing $S_n$ and $N_m$ by the total number of structures and nodes, $pt$ and $qt$, respectively. One then obtains recursively the following solutions for the normalized distributions at statistical equilibrium, $\{\mathcal{S}_n^*\}$ and $\{\mathcal{N}_m^*\}$: 
%\begin{eqnarray}
%\!\!\!\! & S_n^* = \dfrac{\prod _{k=1}^{n-1} k\left(1-p\right)}{\prod _{k=1}^{n} \left(1+k\left(1-p\right)\right)} \label{m3}\\
%\!\!\!\! & N_m^* = \dfrac{\prod _{k=1}^{m-1} k\left(1-q\right) }{\prod _{k=1}^{m} \left(1+k\left(1-q\right) \right)} \label{m4}
%\end{eqnarray}
%\begin{equation}
%S_n^* = \dfrac{\prod _{k=1}^{n-1} k\left(1-p\right)}{\prod _{k=1}^{n} \left[1+k\left(1-p\right)\right]} \; ; \;\; N_m^* = \dfrac{\prod _{k=1}^{m-1} k\left(1-q\right) }{\prod _{k=1}^{m} \left[1+k\left(1-q\right) \right]}
%\label{m4}
%\end{equation}
\begin{eqnarray}
& \mathcal{S}_n^* = \dfrac{\prod _{k=s}^{n-1} k\Omega _s}{\prod _{k=s}^{n} \left(1+k\Omega _s\right)} \;\; \textrm{where } &\;\; \Omega _s = \frac{1-p}{1+p(s-1)} \\
& \mathcal{N}_m^* = \dfrac{\prod _{k=1}^{m-1} k\Gamma _s}{\prod _{k=1}^{m} \left(1+k\Gamma _s\right)} \;\; \textrm{where } &\;\; \Gamma _s = \frac{1+p(s-1)-q}{1+p(s-1)} \; ,
\label{m4}
\end{eqnarray}
which scale as indicated in Table \ref{table}, $\mathcal{N}^*_m \propto m^{-\gamma_N}$ and $\mathcal{S}^*_n \propto n^{-\gamma_S}$.

\begin{table}[h!]
\begin{center}
\begin{tabular}{ @{}l c c@{} } \hline
System base $s$ & Membership scaling $\gamma_N$ & Size scaling $\gamma _S$\\
\hline 
Node ($s=1$) & $\left(2-q\right)/\left(1-q\right) $ & $\left(2-p\right)/\left(1-p\right)$ \\
Link ($s=2$) & $\quad \left[2\left(p+1\right)-q\right]/\left(1+q-p\right) \quad$ & $2/\left(1-p\right)$\\
\hline
\end{tabular}
\end{center}
\vspace{-0.45cm}
\caption{Exponents of the power-law distributions of structures per element (membership) and of elements per structure (size) at statistical equilibrium.  One easily verifies that the membership scaling of link-based systems with $p\!=\!q\!=\!1$ corresponds to that of the BA model ($\gamma _N \!=\! 3$), and that node-based systems with $q=1$ reproduce Simon's model. See Supplemental Material for the derivation \cite{SM}.}
\label{table}
\end{table}

\begin{figure*}[!htb]
  \centering
  \subfigure[]{\includegraphics[width=0.326\linewidth]{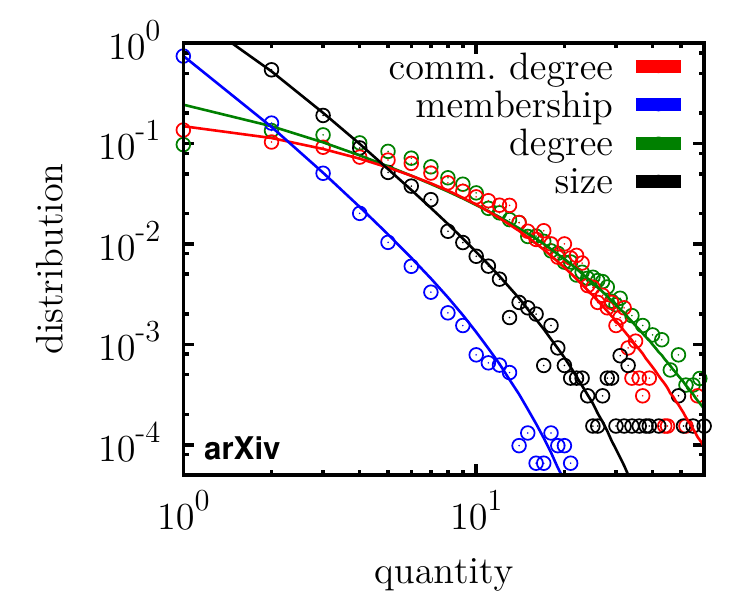} \label{arxiv}}         
  \subfigure[]{\includegraphics[width=0.326\linewidth]{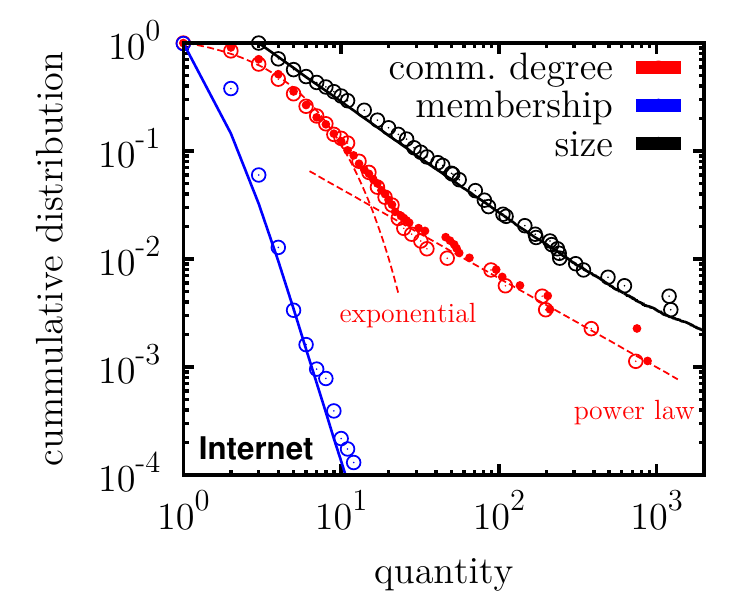} \label{internet}}
  \subfigure[]{\includegraphics[width=0.326\linewidth]{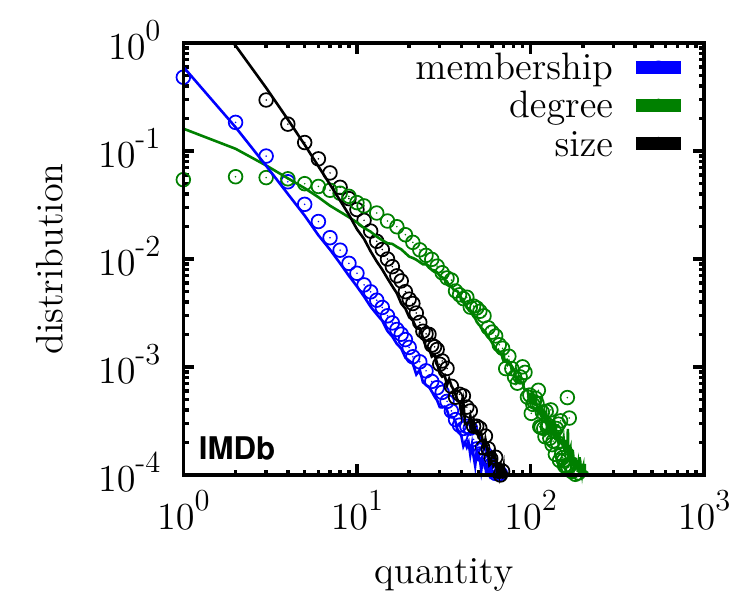} \label{imdb_res}}
  \caption{(Color online) Circles: distributions of topological quantities for (a) the \emph{cond-mat arXiv} circa 2005; (b) Internet at the level of autonomous systems circa 2007; (c) the IMDb network for movies released since 2000. Solid lines: average over multiple realizations of the SPA process with (a) $p = 0.56$ and $q = 0.59$; (b) $p = 0.04$ and $q = 0.66$; (c) $p = 0.47$ and $q = 0.25$. For each realization, iterations are pursued until an equivalent system size is obtained. The Internet data highlights the transition between exponential and scale-free regimes in a typical community degree distribution. It is represented by a single realization of SPA (dots), because averaging masks the transition.}
  \label{results}
\end{figure*}

\paragraph*{\textbf{Results and discussions.}}

There are three distributions of interest which can be directly obtained from SPA: the membership, the community size, and the community degree distributions. In systems such as the size of business firms or word frequencies, these distributions suffice to characterize the organization. To obtain them, the SPA parameters, $q$ and $p$, are fitted to the empirical scaling exponents of the membership and community size distributions. In complex networks, one may also be interested in the degree distribution. Additional assumptions are then needed to determine how nodes are interconnected within communities (specified when required).

The first set of results considered is the community structure of the coautorship network of an electronic preprints archive, the \emph{cond-mat arXiv} circa 2005 [Fig. \ref{arxiv}], whose topology was already characterized using a clique percolation method \cite{palla05}. Here, the communities are detected using the link community algorithm of Ahn \emph{et al.} \cite{ahn}, confirming previous results. 

Using only two parameters, our model can create a system of similar size with an equivalent topology according to the four distributions considered (community sizes, memberships, community degree and node degree). Not only does SPA reproduce the correct density of structures of size 2, 3, 4 or more, but it also correctly predicts \emph{how} these structures are interconnected via their overlap, i.e., the community degree. This is achieved without imposing any constraints whatsoever for this property. The first portion of the community degree distribution is approximately exponential; a behavior which can be observed in other systems, such as the Internet [Fig. \ref{internet}] and both a protein interaction and a word-association network \cite{palla05}. To our knowledge, SPA is the first growth process to reproduce such community structured systems. 

Moreover, assuming fully connected structures, SPA correctly produces a similar behavior in the degree distribution of the nodes. Obtaining this distribution alone previously required two parameters and additional assumptions \cite{barabasi00}. In contrast, SPA shows that this is a signature of a \emph{scale-free community structure}. This is an interesting result in itself, since most observed degree distributions follow a power law only asymptotically. Furthermore, this particular result also illustrates how self-similarity between different structural levels (i.e., node degree and community degree distributions) can emerge from the scale-free organization of communities.

%This study brings a new perspective on the self-similarity of complex networks \cite{song}. Our model shows that it is simply a consequence of preferential attachment at the level of communities: the scale-free organization is inherited by the lower structural levels.

%The cumulative distributions of a second set of results, the Internet at the level of autonomous system circa 2007, is used in Fig. \ref{internet} to highlight the transition between exponential and scale-free regimes in a typical community degree distribution. Once again, SPA is able to generate an equivalent topology.

Finally, the Internet Movie Database coacting network is used to illustrate how, for bigger and sparser communities which cannot be considered fully connected, one can still easily approximate the degree distribution. We first observe that the mean density of links in communities of size $n$ approximately behaves as $\log(n)/n$ (see Supplemental Material \cite{SM}). Then, using a simple binomial approximation to connect the nodes within communities, it is possible to approximate the correct scaling behavior for the degree distribution [Fig. \ref{imdb_res}]. This method takes advantage of the fact that communities are, by definition, homogeneous such that their internal organization can be considered random.

\paragraph*{\textbf{Conclusion and perspective.}}
In this Letter, we have developed a complex network organization model where connections are built through growing communities, whereas past efforts typically tried to arrange random links in a scale-free, modular and/or self-similar manner.  Our model shows that these universal properties are a consequence of preferential attachment at the level of communities: the scale-free organization is inherited by the lower structural levels.

Looking at network organization beyond the link is also useful to account for missing links \cite{clauset} or to help realistic modeling \cite{hebert10, newman10}. For instance, this new paradigm of scale-free community structure suggests that nodes with the most memberships, i.e., structural hubs, are key elements in propagating epidemics on social networks or viruses on the Internet. These structural hubs connect many different neighborhoods, unlike standard hubs whose links can be redundant if shared within a single community.

There is no denying that communities can interact in more complex ways through time \cite{palla07}. Yet, from a statistical point of view, those processes can be neglected in the context of a structurally preferential growth. Similarly, even though other theories generating scale-free designs exist \cite{doyle}, they could also benefit from generalizing their point of view to higher levels of organization.

\paragraph*{\textbf{Acknowledgements}} The authors wish to thank Yong-Yeol Ahn \textit{et al.} for their link community algorithm; Gergely Palla \textit{et al.} for providing the CFinder software and the arXiv data set; Mark Newman for the Internet dataset; and The Internet Movie Database available at www.imdb.com. This research was funded by CIHR, NSERC and FQRNT.

%\bibliographystyle{apsrev}
%\bibliography{./Biblio.bib}
%~\\

\end{document}